\documentclass[aps,prl,twocolumn,groupedaddress]{revtex4}
\usepackage{graphicx}
\usepackage{latexsym}

\begin{document}

\title{What do we learn from the local geometry of glass-forming
  liquids?}

\author{Francis W. Starr$^1$, Srikanth Sastry$^2$, Jack F. Douglas$^1$,
  and Sharon C. Glotzer$^{1,3}$ }

\affiliation{$^1$ Polymers Division and Center for Theoretical and
Computational Materials Science, National Institute of Standards and
Technology, Gaithersburg, Maryland 20899, USA}
\affiliation{$^2$ Jawaharlal Nehru Centre for Advanced Scientific Research, 
Jakkur Campus, Bangalore 560064, INDIA}
\affiliation{$^3$ Departments of Chemical Engineering and Materials
Science and Engineering, University of Michigan, Ann Arbor, MI 48109, USA}

\date{June 7, 2002}
 
\begin{abstract}
  We examine the local geometry of a simulated glass-forming polymer
  melt.  Using the Voronoi construction, we find that the distributions
  of Voronoi volume $P(v_V)$ and asphericity $P(a)$ appear to be
  universal properties of dense liquids, supporting the use of packing
  approaches to understand liquid properties.  We also calculate the
  average free volume $\langle v_f \rangle$ along a path of constant
  density and find that $\langle v_f \rangle$ extrapolates to zero at
  the same temperature $T_0$ that the extrapolated relaxation time
  diverges.  We relate $\langle v_f \rangle$ to the Debye-Waller factor.
\end{abstract}
\bigskip
\pacs{PACS numbers: 61.25.Hq, 61.43.Fs, 61.82.Pv, 64.70.Pf}

\maketitle

A fundamental mystery in the formation of glasses is the relationship of
liquid structure to dynamics.  While it has long been appreciated that
many equilibrium and transport properties of dense fluids depend on the
space available for molecular motion, the lack of methods to accurately
compute or measure ``free volume'' and other measures of local structure
has limited the development of this perspective for understanding fluid
properties.  Rahman~\cite{rahman} suggested that the Voronoi cells
(where a cell is defined as the sub-volume whose interior is closer to a
specific atomic or molecular vertex than to any other vertex) may
provide useful information about the local molecular environment.
Furthermore, Voronoi volume $v_V$ has been considered a possible measure
of the local free volume that might be correlated with liquid
relaxation~\cite{cohen-turnbell,cohen-grest}, and there has been some
tentative evidence to support this proposition~\cite{medvedev,hiwatari}.
Unfortunately, an unequivocal definition of free volume has been
elusive, making it difficult to quantitatively test these ideas.
However, for hard spheres, free volume can be rigorously defined and
quantitatively related to the equation of
state~\cite{hoover,speedy,speedy-reiss}.

In this Letter, we focus on the local geometry of a simulated
glass-forming polymer melt using a Voronoi analysis and a free volume
approach based on a mapping to hard spheres.  From the Voronoi approach,
we find the striking result that the distribution of cell volumes
$P(v_V)$ and the distribution of cell asphericities $P(a)$ are universal
functions of temperature $T$ and density $\rho$ over a wide range of the
liquid state, independent of the interaction potential.  While we are
unable to quantitatively explain the origin of this universality, the
existence of regularity in the structure of liquids is likely connected
to the success of liquid state theories that focus on packing effects
arising from core repulsion~\cite{wca+others}.  Utilizing a free volume
definition related to hard spheres, we find that the extrapolated
average free volume $\langle v_f \rangle$ appears to vanish at the same
temperature $T_0$ where the extrapolated relaxation time $\tau$ of the
coherent intermediate scattering function diverges.  By relating
$\langle v_f \rangle$ to the Debye-Waller (DW) factor $\langle u^2
\rangle$, we find that $\tau \sim \exp (\tilde{u}^2/\langle u^2 \rangle
)$ ($\tilde{u}$ is a constant), as experimentally
observed~\cite{buchenau}, and predicted by several
models~\cite{wolynes-jeppe}.

Our results are based primarily on molecular dynamics simulations of a
melt containing 100 chains of ``bead-spring'' polymers, each chain
consisting of 20 monomers.  All monomers interact via a force-shifted
Lennard Jones potential, truncated at 2.5, in standard reduced units.
Neighboring monomers along a chain also interact via a FENE spring
potential.  The FENE parameters are $k=30$ and $R_0 = 1.5$, and are
chosen to avoid crystallization.  The dynamics of this model (with only
the potential shifted) was studied in ref.~\cite{benneman}.  Most of our
simulations are in the range $0.35<T<4.0$ at constant density $\rho =
1.0$, and as a result the average Voronoi cell volume $\langle v_V
\rangle = 1.0$ is constant.  We also consider several additional
densities to demonstrate the generality of our results.  For reference,
we fit $\tau$ to the power law form $\tau \sim (T/T_{\rm
  MCT}-1)^{-\gamma}$ expected from mode coupling theory
(MCT)~\cite{mct}, and estimate the crossover temperature $T_{\rm MCT} =
0.35 \pm 0.01$, where the uncertainty represents the range of $T$ for
which a power law fits well to the data.

\noindent{\it A. Voronoi Cell Volume and Asphericity}

\begin{figure}
\begin{center}
\includegraphics[clip,width=8.6cm]{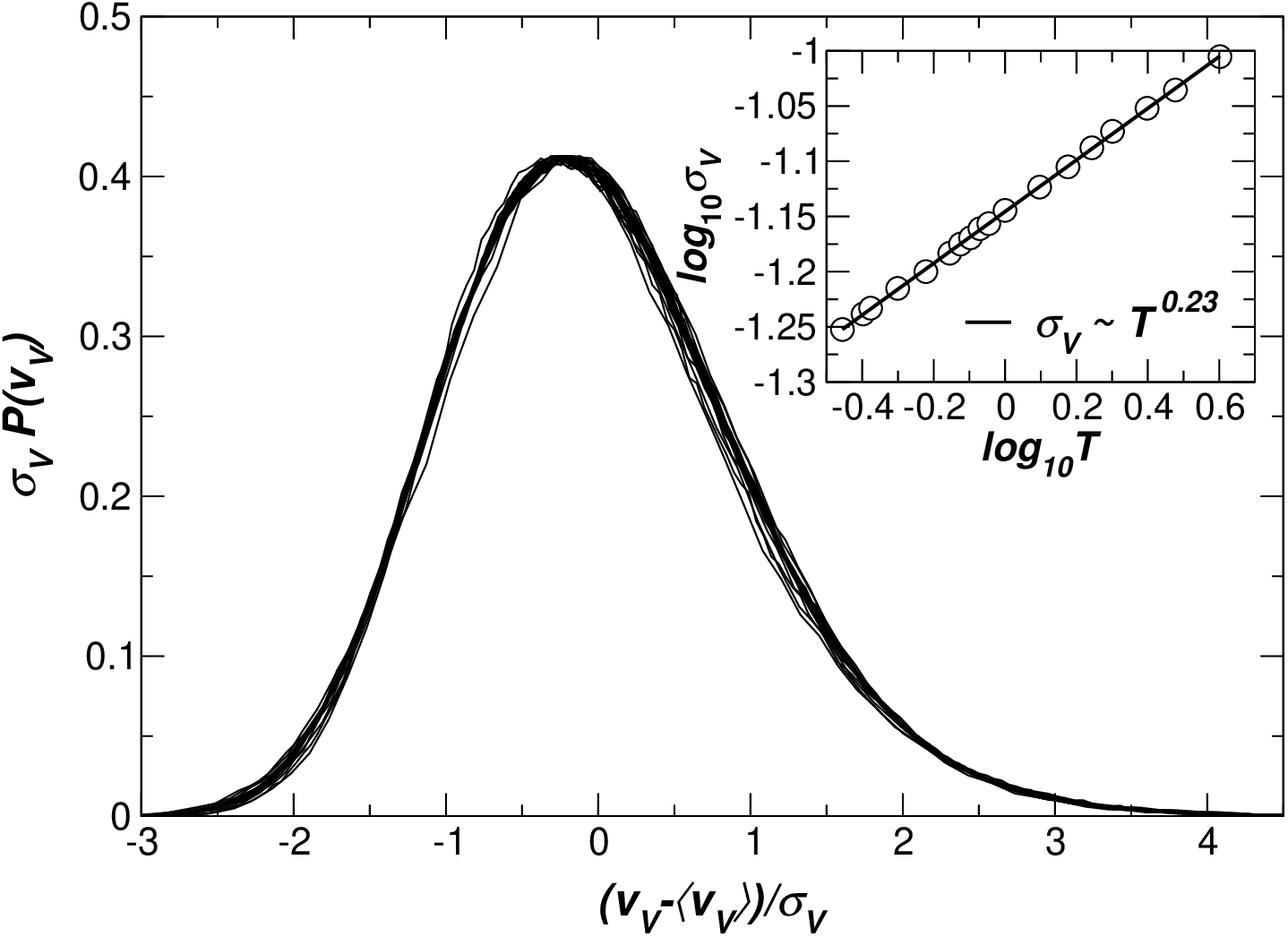}
\end{center}
\caption{Distribution of Voronoi cell volumes, shifted by the average
  voronoi volume $\langle v_V\rangle$ and scaled by the root-mean-square
  deviation $\sigma_v$.  The figure includes data from 21 different
  state points of variable $T$ and $\rho$ (ranging from a very high $T$
  liquid to supercooled states), as well as data from 7 state points of
  simulated liquid water, and one state point of simulated silicon.  Most
  of the spread in the collapse is from the water data, for which we
  have inferior statistics.  The inset shows the temperature dependence
  of $\sigma_v$ along the $\rho = 1.0$ isochore.}
\label{fig:Voronoi-dist}
\end{figure}

The sensitivity of the Voronoi analysis to local structure is
particularly helpful for understanding changes along an isochoric
cooling path, since the changes in local environment are more subtle
than along an isobaric path.  We implement an efficient algorithm for
generating the Voronoi tessalation and the Delauney
simplices~\cite{voronoi-algorithm}.  From these calculations we first
consider the statistical properties of the Voronoi cell volumes $v_V$.
This leads to the striking result that the distribution of Voronoi cell
volumes $P(v_V)$ appears to be universal, where the standard deviation
$\sigma_v^2 \equiv \langle v_V^2 \rangle - \langle v_V \rangle^2$
decreases upon cooling (Fig.~\ref{fig:Voronoi-dist}).  The scaling of
$P(v_V)$ suggests the existence of a single underlying geometrical
structure of the system, and that system specifics, such as temperature,
density, and interaction potential, are absorbed into the average and
variance of the distribution.  To further test this possibility, we
calculate $P(v_V)$ for $\rho = 0.9, 0.95, 1.05$, and $1.1$ at $T=1.0$
and find that $P(v_V)$ for different densities also scales to the same
master curve.  As an even more stringent test, we calculate $P(v_V)$ for
configurations of simulated liquid water~\cite{spce} and silicon; both
liquids are anomalous due to the presence of directional bonding in the
liquid state. We find that $P(v_V)$ for water and silicon collapses to
the {\it same} universal curve as the polymer data, supporting the
existence of a single underlying distribution of Voronoi volumes for
dense liquids.

Such universal behavior suggests that a limit
theorem~\cite{limit-theorem} governs $P(v_V)$ for a wide range of
possible interactions.  The distribution is best fit by a log-normal
distribution, but the deviations of the fit exceed the quality of the
data collapse.  This universal packing behavior likely explains why
models based on short-range packing repulsion can successfully account
for many liquid state properties~\cite{wca+others}.  This may serve as a
point of departure for theoretical investigation of the origin of the
scaling, which is beyond the scope of the present paper.  We also point
out that scaling must break down at sufficiently low density or high
temperature, since particles are randomly distributed in the ideal gas
limit.

\begin{figure}
\begin{center}
\includegraphics[clip,width=8.6cm]{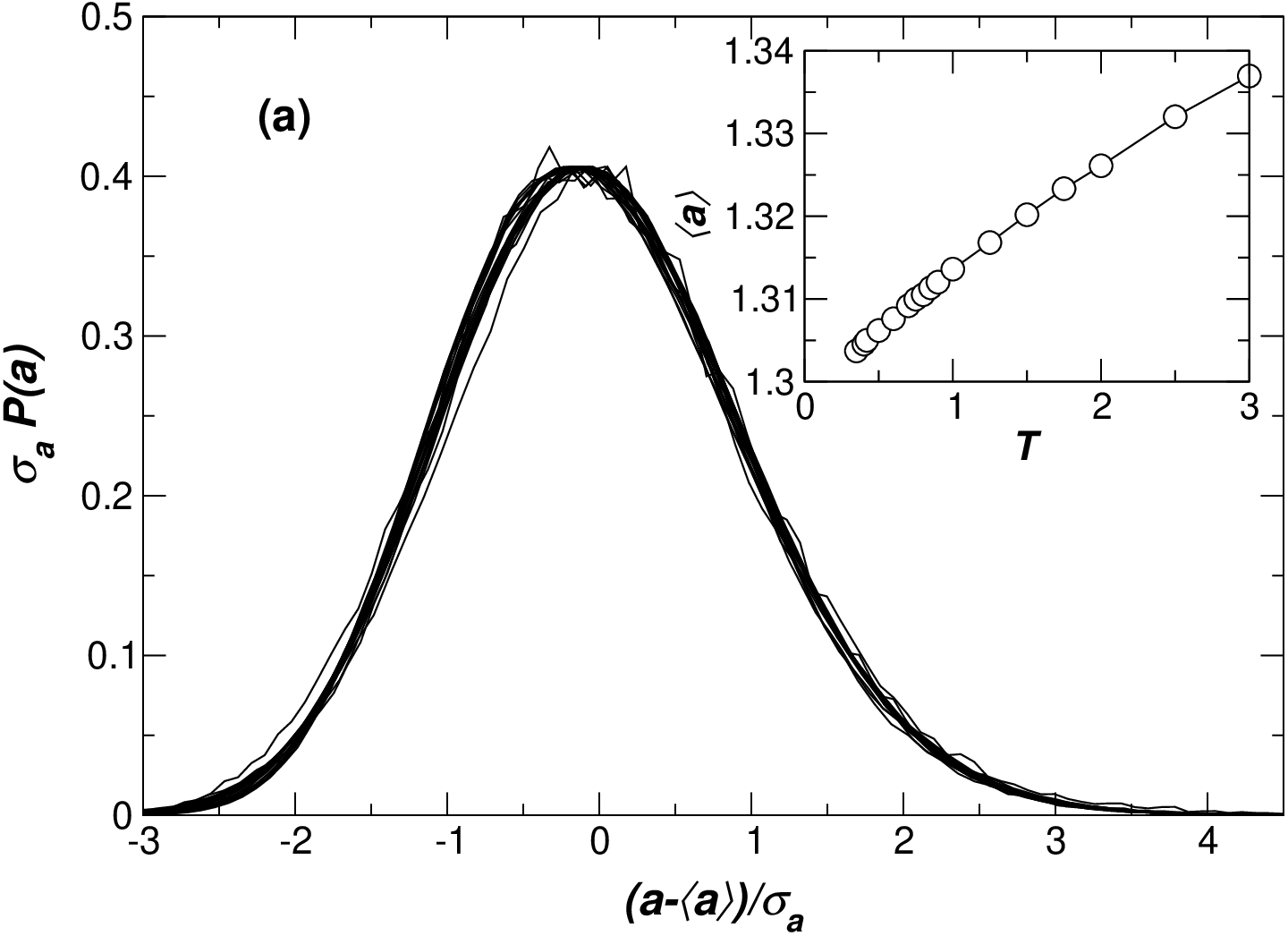}
\end{center}
\caption{Distribution of Voronoi cell asphericities, shifted by the average
  asphericity $\langle a\rangle$ and scaled by the root-mean-square
  deviation $\sigma_a$.  The figure includes data from the same state
  points as Fig.~\ref{fig:Voronoi-dist}.  The inset shows the $T$
  dependence of $\langle a\rangle$ along the $\rho = 1.0$ isochore. }
\label{fig:asphericity}
\end{figure}

The scaling parameter $\sigma_v$ follows a power law $\sigma_v \sim
T^{0.23 \pm 0.01}$ (inset of Fig.~\ref{fig:Voronoi-dist}).  Since
$\sigma_v$ is the fluctuation in Voronoi volume, and compressibility
$\kappa_T$ is a measure of density fluctuations, it is tempting to
relate the quantities~\cite{compressibility}.  However, $\sigma_v$
measures volume fluctuations the size scale of single particles, too
small to expect a fluctuation relation of the form
$\kappa_T=\sigma_v^2/(v_Vk_BT)$ to hold~\cite{compress-note}.  On the
other hand, $\sigma_v$ should be related to the local restoring force of
the fluid felt by a particle, since this controls the susceptibility of
the fluid to local fluctuations.  The restoring force is quantified by
$\Omega_0$, the first non-trivial coefficient in an expansion of the
velocity auto-correlation function~\cite{boon-yip}.  Indeed, it has been
found that $\Omega_0$ scales approximately as $T^{1/4}$~\cite{boon-yip},
which we also confirmed for our system.  This suggests a connection
between $\sigma_v$ and vibrational dynamics (given the uncertainties in
the exponent values).  At fixed $T=1$, we further observe $\sigma_v \sim
\rho^{-3.4}$ over the narrow range $0.9 < \rho < 1.1$.  In the cases of
simulated water and silicon, our data is not able to confirm or exclude
power-law scaling of $\sigma_v$.

We further characterize local geometry by the asphericity $a$ of the
cells, which we define by the ratio of the radius of a sphere with the
same volume as the cell to the distance between the Voronoi vertex and
the nearest cell face.  If the cell is spherical, the ratio is one,
while for any non-spherical cell, the ratio is greater than one.  On
cooling, $a$ decreases by $\approx 3$\% over the range of our
simulations, and thus Voronoi cells become slightly more spherical on
average (inset of Fig.~\ref{fig:asphericity}), as would be expected if
the packing becomes more regular on cooling.  The distribution of
asphericities $P(a)$, like $P(v_V)$, also scales onto a single master
curve using the same scaling method (Fig~\ref{fig:asphericity}).  The
width of the distribution, quantified by the standard deviation
$\sigma_a$, monotonically decreases with decreasing $T$ (not shown),
also expected if packing becomes more regular on cooling.  We find that
$\sigma_a$ follows a power law whose exponent changes slightly at
$T\approx 1$, roughly the ``onset'' temperature for slow
dynamics~\cite{sri-onset} and spatially heterogeneous monomer
motion~\cite{glotzer}.  However, the lack of corroborating evidence from
the simulations of water and silicon suggests that the result for the
polymer system is merely coincidental.  Similar calculations for other
simple and molecular liquids will be useful to identify the common
features of cell geometry in liquids.

\noindent{\it B. Free Volume and Dynamics}

We next focus our attention on possible connections between local
structure and dynamics by examining the free volume.  There have been
numerous definitions of free volume~\cite{cohen-grest,hiwatari,hoover},
and thus it is not clear what the most appropriate and/or useful
definition is.  However, in the case of hard spheres, $v_f$ of a
particle can be unambiguously defined as the volume over which the
center of a sphere can translate, given that all other spheres in the
system are fixed (see, e.g. Fig.~1 of ref.~\cite{sri}, where an
algorithm for the calculation can also be found).  This ``rattle'' free
volume is rigorously related to the equation of state for hard spheres,
as well as the void volume~\cite{speedy-reiss}, and thus is an
attractive definition.

In the case of soft-core repulsion, such a definition of $v_f$ is not
straightforward, since the average distance of closest approach depends
on $T$.  As done in ref.~\cite{ssphere-radius} we define a $T$ dependent
effective hard sphere diameter, or Boltzmann diameter $\sigma_B$, by
$U(r=\sigma_B) = E_c$, where $U$ is the pair potential and $E_c$ is the
average kinetic energy of collision of an isolated pair of particles.
Since $E_c$ for this system is not known, we estimate $E_c = 2k_BT$, the
exact result for hard spheres.  Additionally, we use only the non-bonded
(Lennard-Jones) part of the potential to determine $\sigma_B$, since the
bonding term has little effect on the distance of closest approach.  By
using the $T$ dependent definition of $\sigma_B$, it is possible that
$\langle v_f \rangle$ vanishes along paths of constant density, such as
we study here, in contrast with free volume definitions tied to the
macroscopic density.

\begin{figure}
\begin{center}
\includegraphics[clip,width=8.6cm]{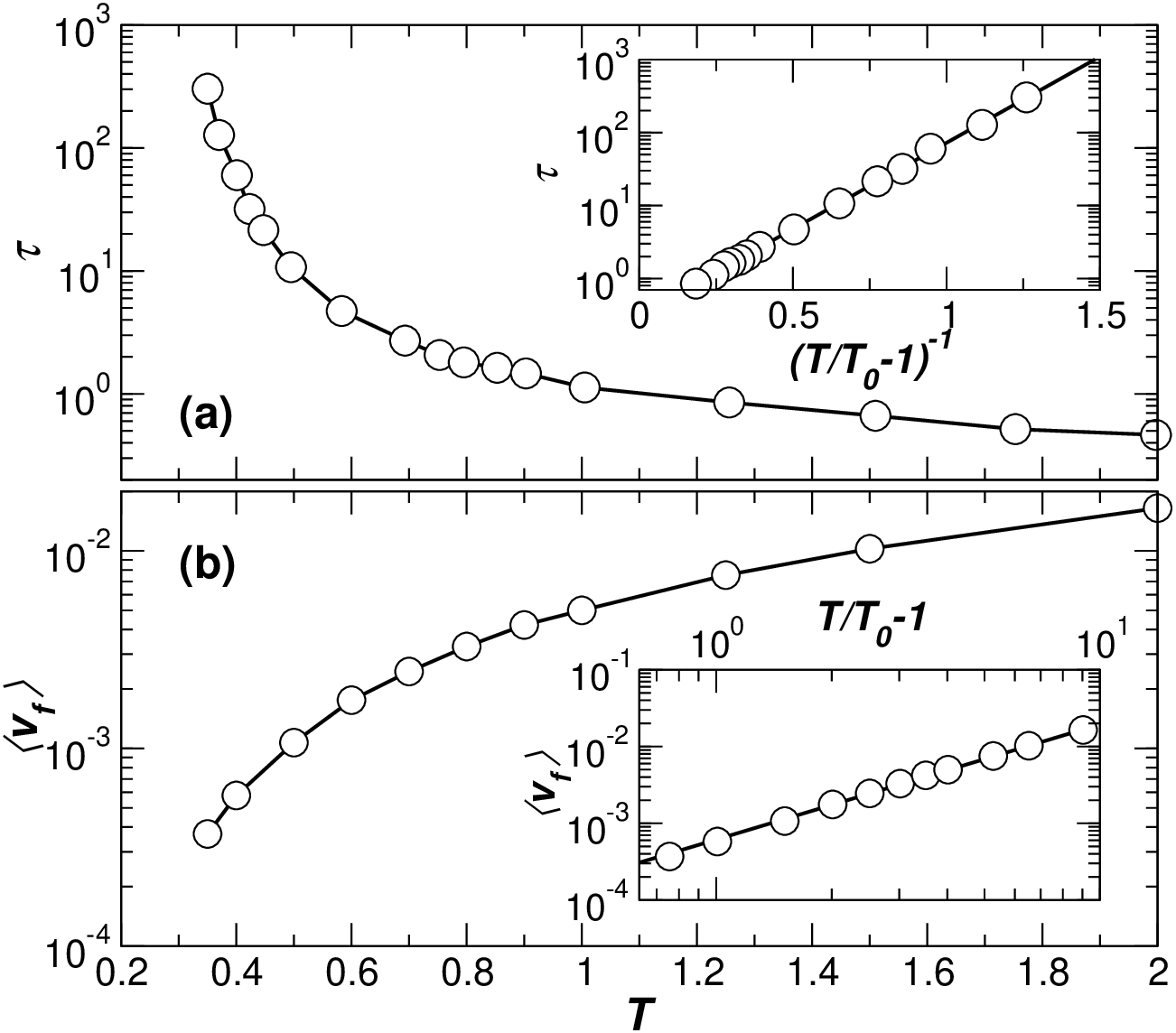}
\end{center}
\caption{(a) The relaxation time $\tau$ of the intermediate scattering
  function $F(q,t)$; we define $\tau$ at the time at which $F(q,t)$ has
  decayed by a factor of $e^{-1}$ at the $q$-value corresponding to the
  first maximum in $S(q)$.  The inset shows the quality of the fit to
  the VFT form.  (b) Average free volume per particle $\langle v_f
  \rangle$.  The inset is a log-log plot to emphasize the power law
  behavior }
\label{fig:tau-freevol}
\end{figure}

We first consider the $T$ dependence of the average free volume per
monomer $\langle v_f \rangle$ and $\tau$ (Fig.~\ref{fig:tau-freevol}).
As expected, $\langle v_f \rangle$ decreases on cooling while $\tau$
increases.  More significantly, we find that $\langle v_f \rangle$ may
be well approximated by a power law $\langle v_f \rangle \sim
(T/T_0-1)^{\eta}$, where $T_0 = 0.22$ and $\eta = 1.46$, if we allow
both parameters to be free.  For reasons that we will discuss, we expect
$\eta = 3/2$, which results in a best fit value of $T_0 = 0.20$.  We fit
$\tau$ to the Vogel-Fulcher-Tammann (VFT) equation $\tau \sim \exp
[A/(T-T_0)]$ and independently obtain an extrapolated divergence
temperature $T_0 = 0.20$, which is typically slightly below the
laboratory defined $T_g$.  Hence, $\tau$ seems to diverge at the same
$T$ at which $\langle v_f \rangle$ extrapolates to zero, consistent with
the possibility that the glass transition is related to vanishing of
free volume, a central tenet of most free volume
approaches~\cite{vanish-note}.  By eliminating the $T$ dependence from
the expressions for $\langle v_f \rangle$ and $\tau$, we obtain the
parametric relation
\begin{equation}
\tau \sim \exp[(\tilde{v}/\langle v_f \rangle )^{2/3}]
\label{eq:tau-fv}
\end{equation}
where $\tilde{v}$ is a constant.  This relation is similar to, but
distinct from the Doolittle expression $\tau \sim \exp(v_0/v_f)$, later
rationalized by the Cohen-Turnbull models that assume an exponential
distribution of free volume~\cite{cohen-turnbell}.

\begin{figure}[t]
\begin{center}
\includegraphics[clip,width=8.6cm]{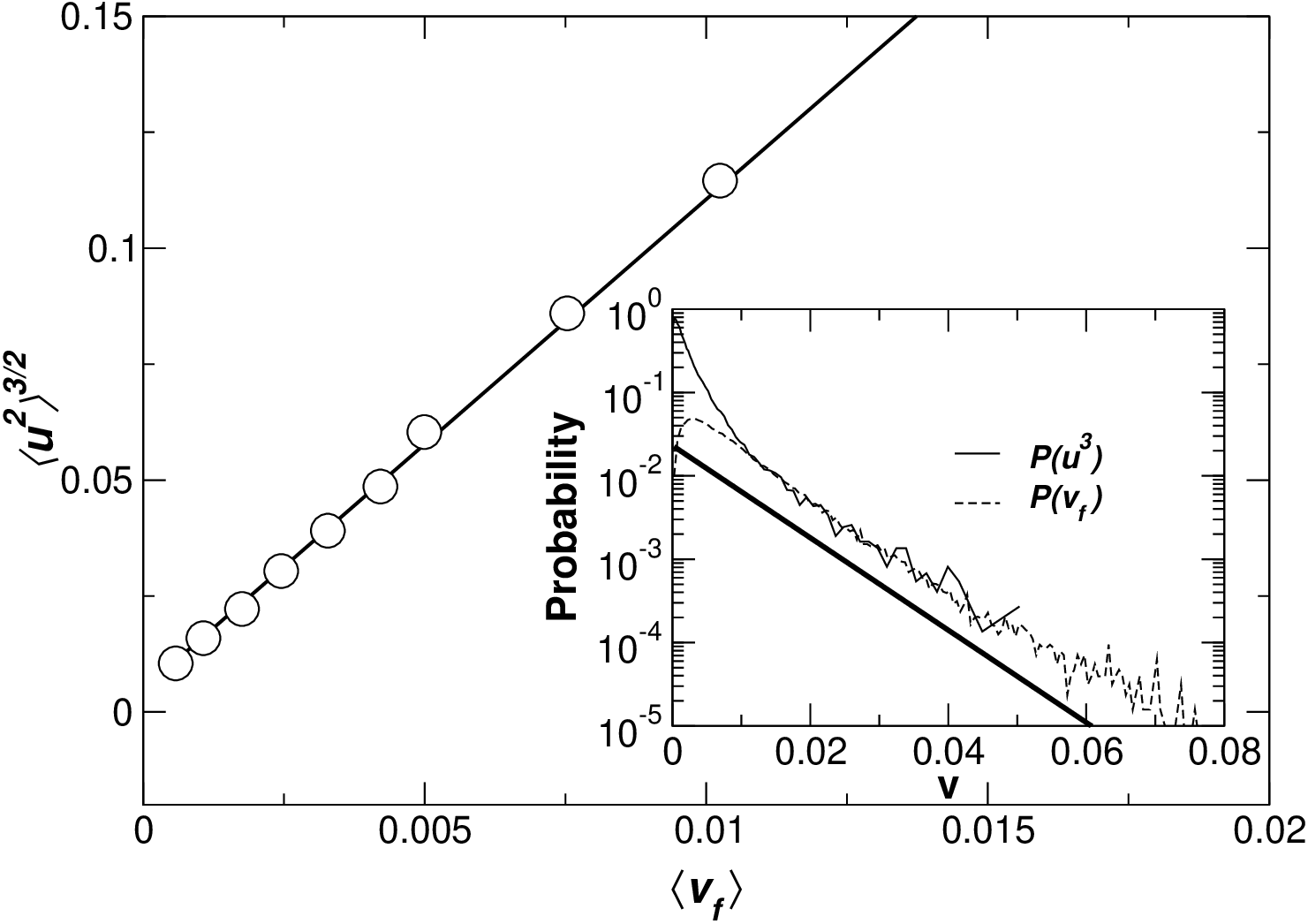}
\end{center}
\caption{Parametric plot of the Debye-Waller factor $\langle u^2
  \rangle$ raised the the $3/2$ power (such that is has units of volume)
  as a function $\langle v_f \rangle$.  The line indicates a
  least-square fit.  The inset shows the distributions $P(u^3)$ and
  $P(v_f)$.  The data for $P(u^3)$ is arbitrarily shifted to emphasize
  the similarity of the distributions.  The bold line in an exponential
  $\exp (v/0.2^3)$, plotted as a guide to the eye. }
\label{fig:cage-freevol}
\end{figure}

In order to explain the power-law dependence of $\langle v_f \rangle$ on
$T$, and hence the unexpected relation between $\tau$ and $\langle v_f
\rangle$, as well as provide an experimental connection, we must better
understand the physical origin of $\langle v_f \rangle$.  Since the free
volume measures the space over which a particle can move before
encountering the exclusion volume of neighboring particles (in the hard
sphere mapping), it is natural to expect that free volume and the
Debye-Waller factor, a measure of cage size, might be related.  To test
this, we define the DW factor $\langle u^2 \rangle \equiv \langle
r^2(t=1.022) \rangle$, where $t = 1.022$ is the approximate time of the
crossover from ballistic to caged motion of the mean-squared
displacement $\langle r^2(t) \rangle$.  In Fig.~\ref{fig:cage-freevol}
we make a parametric plot of $\langle v_f \rangle$ and $\langle
u^2\rangle ^{3/2}$ and find a linear proportionality, supporting the
hypothesis that free volume should be related to the DW factor.  As a
more stringent test, we consider the distributions $P(v_f)$ and
$P(u^3)$, where $P(u^3)$ is calculated in the same fashion as the van
Hove correlation function.  The inset of Fig.~\ref{fig:cage-freevol}
shows that both distributions are nearly exponential, but that
stretching occurs at large volume.  Exponential decay of $P(v_f)$ at
large $v_f$ is essential to recover the Doolittle relation in the
Cohen-Turnbull formulation of free volume theory; since the Doolittle
relation does not appear to hold for our data, the significance of the
deviation from exponential decay is unclear.  More importantly, the
similarity of $P(v_f)$ and $P(u^3)$ for sufficiently large volume allows
us to interpret $v_f$ as a measure of the DW factor, and apply ideas
that have already been developed relating $\langle u^2 \rangle$ and bulk
relaxation properties.

We substitute $\langle u^2 \rangle^{3/2}$ for $\langle v_f \rangle$ in
Eq.~\ref{eq:tau-fv} and obtain
\begin{equation}
\tau \sim \exp (\tilde{u}^2/\langle u^2\rangle ).
\label{eq:tau-r2}
\end{equation}
Several arguments have been put forth for this phenomenologically
observed relation~\cite{buchenau}, based on the idea that the effective
force constant localizing a particle is inversely proportional to
$\langle u^2 \rangle$ and directly proportional to the energy barrier
height~\cite{wolynes-jeppe}.  The $T$ dependence of $\langle u^2
\rangle$ is controlled by two factors: (i) inertial energy, resulting in
a term proportional to $k_BT$, counterbalanced by (ii) a nearly $T$
independent elastic energy proportional to $G_\infty \delta^3$, where
$G_\infty$ is the shear modulus and $\delta^3 \sim \mathcal{O}(v_V)$.
This immediately yields $\langle v_f \rangle \sim (T/T_0-1)^{3/2}$,
where $T_0 \propto G_\infty \delta^3/k_B$, since $\langle v_f \rangle
\propto \langle u^2 \rangle^{3/2}$.  Such a relationship is consistent
with the VFT expression, as demonstrated by simple substitution.
Additionally, the Adam-Gibbs expression $\tau \sim \exp (A/TS_{\rm
  conf})$, successfully applied to a variety of model
liquids~\cite{AG-success}, suggests a non-trivial relationship to the
configurational entropy $S_{\rm conf}$; specifically, $TS_{\rm conf}
\propto \langle u^2\rangle$ or $TS_{\rm conf} \propto \langle v_f
\rangle^{2/3}$.  This offers an area of future consideration.

We have focused our attention on a limited subset of the predictions of
free volume theories.  We did not find support for several additional
hypotheses.  Specifically, we also find that (i) free volume percolates
at $T$ well above $T_g$ and (ii) that that no significant correlation
exists between local volume of a specific monomer and its mobility.
More details on these results will be provided in a future publication.

We thank J. Baschnagel, W. Kob, F. Sciortino, and V. Novikov for helpful
discussions.


\begin{thebibliography}{100}
\bibitem{compress-note} We check the possible relation between
  $\sigma_v^2$ and $\kappa_T = 1/\rho (d\rho/dP)_T$ by numerical
  differentiation of $(P,\rho)$ isotherms.  We find that $\kappa_T$
  approximately scales as $T^{0.54}$.
\bibitem{vanish-note} When $\langle v_f \rangle$ of a system is
  extremely small, we expect that additional factors --- such as the
  topology of the free volume --- become important.  However the
  extrapolation to the vanishing $T$ of $\langle v_f \rangle$ should
  correlate well with the vitrification temperature.
\end{thebibliography}
\end{document}